\documentclass[floatfix,showpacs,11pt,nofootinbib]{revtex4}
\usepackage{amssymb,amsmath}
\usepackage{graphicx,float}
\usepackage{indentfirst}

\newcommand{\beq}{\begin{equation}}
\newcommand{\eeq}{\end{equation}}
\newcommand{\bea}{\begin{eqnarray}}
\newcommand{\eea}{\end{eqnarray}}

\def\({\left(}
\def\){\right)}

\begin{document}

\title{Smooth braneworld models possibility in modified gravities}
\author{G. P. de Brito}
\email{gustavopazzini@gmail.com}
\affiliation{ Centro Brasileiro de Pesquisas F\'{\i}sicas (CBPF),
Rua Dr. Xavier Sigaud 150, Urca, 22290-180, Rio de Janeiro, RJ, Brazil.}
\author{J. M. Hoff da Silva}
\email{hoff@feg.unesp.br}
\author{P. Michel L. T. da Silva}
\email{pmichel@fc.unesp.br}
\author{A. de Souza Dutra}
\email{dutra@feg.unesp.br}
\affiliation{Departamento de F\'{\i}sica e Qu\'{\i}mica, Universidade Estadual Paulista,
Av. Dr. Ariberto Pereira da Cunha, 333, Guaratinguet\'{a}, SP, Brazil.}

\begin{abstract}
It is shown that the consideration of the braneworld consistency conditions
within the framework of bulk modified gravities allows for the existence of thick branes in the five-dimensional case with compact extra dimension. In studing the specific consistency conditions in the Brans-Dicke gravity we were able to show that
the brane generating scalar field potential is relevant for relaxing the
gravitational constraints.
\end{abstract}

\maketitle


\section{Introduction}

The idea that we live in a four-dimensional noncompact Universe (the brane)
embedded in a higher dimensional spacetime have received considerable amount
of attention in the literature (for some comprehensive reviews see refs. 
\cite{Csaki,Gabadadze,Sundrum,Rubakov}). In fact, part of this attention to
braneworld scenarios was motivated by the possibility of solving certain
problems in particle physics, for instance, the so-called mass hierarchy
problem. The Randall-Sundrum (RS) framework provides a very beautiful and
simple approach to this problem \cite{RS1}. In that context, the small ratio
between the eletroweak scale and the Planck scale appears as a consequence
of an exponential suppression in the Higgs mass term, due to gravitational
effects in a higher dimensional spacetime endowed with warped geometry. 
\newline
\indent In the RS model there are two 3-branes (our 3+1 dimensional
Universe), with opposite tensions, localized at the ends of an $S^{1}/%
\mathbf{Z}_{2}$ orbifold. In this framework, the branes appear as singular
sources in the five-dimensional Einstein field equations. Since then,
several extensions of the RS model were proposed. A very interesting
approach was that adopted by De Wolfe \textit{et.al.} \cite{DEWOLFE}, and independently, by Gremm \cite{Gremm}, where the full spacetime is
a five-dimensional manifold with warped geometry and, the brane-bulk
structure is generated by a scalar field coupled to gravity. In this
context, the brane is interpreted as being a domain-wall where the standard
model fields are localized in.\newline
\indent It is quite difficult, however, to make any judgment regarding the
consistency of many models. A special attention must be taken to the
question of whether the spacetime geometry, along with background fields and
branes, solve the set of Einstein field equations. In a seminal paper,
Gibbons, Kallosh and Linde derived a set of consistency conditions \cite%
{Gibbons}, the so-called sum rules, which are helpful to solve the
consistency problem, without solving the dynamical equations. Particularly,
one important conclusion obtained in Ref. \cite{Gibbons} is that braneworld
scenarios with compact extra dimensions demands the inclusion of negative
tension branes, which is a disappointing result, since such objects are
gravitationally unstable. Some time later, Leblond, Myers and Winters
generalized the brane world sum rules to an arbitrary number of dimensions 
\cite{Leblond}. These authors also shown that it is possible to evade the
necessity of negative tension brane by increasing the number of dimensions
of the compact internal space.\newline
\indent An important result derived from the sum rules \cite{Gibbons} was
the no-go theorem regarding the impossibility of construction of smooth
braneworld scenarios with compact internal spaces. The no-go theorem may be
stated as follows: smooth generalizations of the RS scenario without
singular sources are inconsistent with compact extra dimension. This
conclusion was based in the following result 
\begin{equation}
\oint \Phi ^{\prime }\cdot \Phi ^{\prime }=0,  \label{no-go}
\end{equation}%
where $\Phi $ is a scalar field, supposedly used as source for the branes,
and prime means derivative with respect to the extra dimension. Note that
the only possible way to satisfy the above equations is $\Phi $ taking constant
and, as a consequence, there is no brane-bulk structure.\newline
\indent The main aim of this paper is to show that it is possible to escape from
the no-go theorem of the last paragraph using modified gravities. In the
last year, Ahmed and Grzadkowski attempted to evade this theorem by using a
non-minimal scalar-gravity coupling, nevertheless the result was negative 
\cite{Ahmed}. In fact, a crucial diference between the approach used in Ref. \cite{Ahmed} and the presented here is that in the former case the non-minimal coupled scalar field is, at the same time, the scalar field responsible to generate the brane. Therefore, when the gravitational system of equations are to be satisfied along with the $Z_2$ symmetry, there are to much constraints over the very same field, and it turns out that only the trivial shape is allowed. In contrast, in our case even when the gravity sector is also implemented by a scalar field, there is another scalar field shaping the brane.

A clue on the use of modified gravities can be read from the
net result of Ref. \cite{OT}. In Ref. \cite{OT} it was shown that the presence of a Gauss-Bonnet term in the five-dimensional lagrangian can be used to, in some cases, turn the braneworld sum rules less severe. On the other hand the functional form of the resulting consistency conditions in \cite{OT} is, perhaps, too much complicated, encouraging the searching for different solutions.

In the present work, we revisited the sum rules in
the context of $f(R)$ and Brans-Dicke (BD) gravities showing that in both
cases the no-go theorem of the previous paragraph may be relaxed, allowing
for the construction of smooth versions of braneworld scenarios in
accordance with compact internal spaces. In a sense, given the equivalence
of $f(R)$ and Brans-Dicke theory in some specific cases, the possibility of
circumventing the no-go theorem in both theories is expected. On the other
hand, as stated, since the equivalence is only valid for specific cases, a
full treatment of both cases is necessary. It is interesting to note that
the consistency conditions were already incorporated in the context of $f(R)$
\cite{Hoff1} and BD gravities \cite{Hoff2,Hoff3}, in these works it was shown that both $%
f(R)$ and BD theories are suitable to accommodate consistent scenarios
without negative tension branes.

This work is organized as follows: in the next Section we attempt to
investigate the sum rules when applied to the $f(R)$ and Brans-Dicke cases.
As we shall see, both cases allow for existence of smooth $3$-branes in five
dimensions. In the last Section we conclude.

\section{Sum rules for braneworld scenarios}

By considering the spacetime as a $D$-dimensional manifold endowed with a
nonfactorizable geometry, we write the line element as 
\begin{equation}
ds^{2}=G_{AB}(X)dX^{A}dX^{B}=W^{2}(r)g_{\mu \nu }(x)dx^{\mu }dx^{\nu
}+g_{mn}(r)dr^{m}dr^{n},
\end{equation}%
where $W^{2}(r)$ is the warp factor, $X^{A}$ denotes the coordinates of the
full $D$-dimensional spacetime, $x^{\mu }$ stands for the $(p+1)$
coordinates of the noncompact spacetime (brane), and $r^{m}$ labels the $%
(D-p-1)$ directions in the internal compact space. The relevant classical
action for our purposes takes into account the spacetime dynamics coupled to
a scalar field, namely 
\begin{equation}
S=S_{gravity}+\int d^{D}X\,\sqrt{-G}\left( -\frac{1}{2}\partial _{A}\Phi
\partial ^{A}\Phi -V(\Phi )\right) ,
\end{equation}%
where we assume that the scalar field has only dependence on the internal
space coordinates $\Phi =\Phi (r^{m})$. From the above action we obtain the
following expressions for the energy-momentum tensor associated with the
scalar field source 
\begin{equation}
T_{\mu \nu }=-W^{2}g_{\mu \nu }\left( \frac{1}{2}\nabla \Phi \cdot \nabla
\Phi +V(\Phi )\right) ,  \label{energy_momentum_1}
\end{equation}%
and 
\begin{equation}
T_{mn}=\nabla _{m}\Phi \nabla _{n}\Phi -g_{mn}\left( \frac{1}{2}\nabla \Phi
\cdot \nabla \Phi +V(\Phi )\right) .  \label{energy_momentum_2}
\end{equation}%
\indent The key point concerning this construction is to relate $D$%
-dimensional geometric objects with the corresponding geometrical objects on
the brane and on the internal compact space. It can be seen \cite%
{Gibbons,Leblond} that if $\bar{R}$ is the scalar of curvature derived from $%
g_{\mu \nu }$ and $\tilde{R}$ its counterpart in the internal space, then
the following relation holds 
\begin{equation}
\nabla \cdot (W^{\alpha }\nabla W)=\frac{W^{\alpha +1}}{p(p+1)}\left[ \alpha
\,\left( W^{-2}\bar{R}-R_{\mu }^{\mu }\right) +(p-\alpha )\left( \tilde{R}%
-R_{m}^{m}\right) \right] ,  \label{consistency2}
\end{equation}%
where $R_{\mu }^{\mu }=W^{-2}g^{\mu \nu }R_{\mu \nu }$ and $%
R_{m}^{m}=g^{mn}R_{mn}$ are the partial traces such that $R=R_{\mu }^{\mu
}+R_{m}^{m}$ and $\alpha $ is an arbitrary parameter. It must be emphasized
that so far we have not specified anything regarding the gravitational
theory that we are considering. This physical piece is implemented as far as
we choose a dynamical gravitational equation relating the partial traces to
the energy-momentum tensor. Another relevant point is that the left hand
side of Eq. (\ref{consistency2}) vanish upon integration over the internal
compact space.

\subsection{Sum rules in $f(R)$ theories}

In the case where the bulk is gravity described by $f(R)$ theories, its
dynamical equation is given by \cite{review_fR} 
\begin{equation}
F(R)R_{AB}-\frac{1}{2}G_{AB}f(R)+G_{AB}\Box F(R)-\nabla _{A}\nabla
_{B}F(R)=8\pi G_{D}T_{AB},  \label{fR}
\end{equation}%
where $F(R)=df(R)/dR$. Eq. (\ref{fR}) can also be recasted in the equivalent
form below 
\begin{equation}
R_{AB}=\frac{1}{F(R)}\Bigg[8\pi G_{D}\Bigg(T_{AB}-\frac{G_{AB}}{D-2}T\Bigg)%
+\nabla _{A}\nabla _{B}F(R)+\frac{G_{AB}}{D-2}\bigg(\Box F(R)-f(R)+RF(R)%
\bigg)\Bigg],  \label{PT}
\end{equation}%
from which the partial traces are readily computed. Part of the algebraic
manipulation hence forward was performed before \cite{Hoff1}, therefore we
shall only enumerate the main steps.

Computing the partial traces from Eq. (\ref{PT}), inserting the result in (%
\ref{consistency2}), taking into account $T^{\mu}_{\mu}$ and $T^{m}_{m}$
coming from Eq. (\ref{energy_momentum_2}) and, finally, considering the most
appealing case, $D=5$ and $p=3$, we have 
\begin{eqnarray}
\alpha \bar{R} \oint W^{\alpha - 1} = 8 \pi G_5 \oint \frac{W^{\alpha + 1}}{%
F(R)} \bigg[ (3 - \alpha) \Phi^{\prime }\cdot \Phi^{\prime }+ 2(\alpha +
1)V(\Phi) \bigg] + 4 \oint \frac{W^{\alpha + 1} \nabla^2 F(R)}{F(R)} + 
\notag \\
+ \, (\alpha + 1)\oint W^{\alpha + 1} \bigg( R - \frac{f(R)}{F(R)} \bigg) +
(2\alpha + 1) \oint \frac{W^{\alpha + 1} \nabla^\mu \nabla_\mu F(R)}{F(R)}.
\end{eqnarray}
As expected, the General Relativity $\alpha$-family of consistency
conditions can be recovered from the above expression. The most sharp
consistency condition regarding the possibility of smooth branes is given
when $\alpha=-1$, since it eliminates the contribution coming from the
scalar field potential. Hence, setting $\alpha=-1$ and $\bar{R}=0$ (in order
to reproduce the observational evidence of a flat four-dimensional universe)
we have 
\begin{equation}
32 \pi G_5 \oint \frac{\Phi^{\prime }\cdot \Phi^{\prime }}{F(R)} + 4 \oint 
\frac{\nabla^2 F(R)}{F(R)} - \oint \frac{\nabla^\mu \nabla_\mu F(R)}{F(R)}=0.
\label{fina}
\end{equation}
Taking into account that in the present case the bulk scalar of curvature
reads 
\begin{equation}
R = - \frac{4}{W}\nabla^2 W - \frac{\nabla^2 W^{4}}{W^{4}},
\end{equation}
we have, as a consequence, $f(R) \rightarrow f(W)$, $F(R) \rightarrow F(W)$
and $\nabla_{\mu}F(R) = \nabla_{\mu}F(W) = 0$. Thus, Eq. (\ref{fina})
reduces to 
\begin{equation}
\oint \frac{\Phi^{\prime }\cdot \Phi^{\prime }}{F(W)} + \frac{1}{8 \pi G_5}
\oint \frac{\nabla^2 F(W)}{F(W)} = 0.  \label{frelevant}
\end{equation}

The above result is very interesting, since it provides a route to escape
from the no-go theorem mentioned earlier. The inconsistency appearing in Eq.
(\ref{no-go}), straightforwardly recovered from the equation above, relies
on the fact that the left hand side should be positive\footnote{%
At least for solutions with extra dimensional dependence, which are the ones
of our interest.}. However, as we can see, in the context of $f(R)$
theories, there is another resulting term in the relevant sum rule, allowing
for the possibility of smooth braneworld scenarios along with a compact
internal space.

\subsection{The Brans-Dicke case}

As previously stated, the equivalence between $f(R)$ and scalar-tensorial
theories is ensured only for very particular choices of the Brans-Dicke
parameter. Therefore, in order to investigate whether the no-go theorem is
valid in the Brans-Dicke case, we have to adequate our previous approach to
this case \cite{Hoff2,Hoff3}.

The Brans-Dicke dynamics is described by the following equations \cite{scalar,brans}

\begin{eqnarray}
R_{AB}-\frac{1}{2}G_{AB}R &=&\frac{8\pi }{\phi }T_{AB}+\dfrac{\omega }{\phi
^{2}}\left( \nabla _{A}\phi \nabla _{B}\phi -\frac{1}{2}\nabla _{C}\phi
\nabla^{C}\phi G_{AB}\right)  \notag \\
&&+\frac{1}{\phi }\left( \nabla _{A}\nabla _{B}\phi -\frac{8\pi }{%
(D-1)+(D-2)\omega }TG_{AB}\right)  \label{b2}
\end{eqnarray}%
and%
\begin{equation}
\square \phi =-\dfrac{\phi }{2\omega }R+\dfrac{1}{2\phi }\nabla^{A}\phi
\nabla_{A}\phi,  \label{b3}
\end{equation}%
where $T_{AB}$ is the matter stress-tensor and $\omega$ is the so-called
Brans-Dicke parameter. We reinforce that $\phi$ is the Brans-Dicke scalar
field, not to be confused with the kink shape scalar field which generates
the brane. Taking Eqs. (\ref{b2}) and (\ref{b3}) together it is possible to
compute the partial traces for this case. In this vein, Eq. (\ref%
{consistency2}) for the Brans-Dicke theory reads 
\begin{eqnarray}
\nabla \cdot (W^{\alpha }\nabla W)&=&\left.\frac{W^{\alpha +1}}{p(p+1)}%
\Bigg\{ \alpha W^{-2}\overset{\_}{R}+(p-\alpha )\tilde{R}+\frac{8\pi}{\phi}%
\frac{1}{[(D-1)+(D-2)\omega ]}\right.  \notag \\
&\times& \left.\Bigg( T_{\mu }^{\mu}\bigg[(D-p-2)(p-2\alpha)-\alpha
(D-p-3)\omega +(p-\alpha )(D-p-1)\omega \bigg]\right.  \notag \\
&+& \left. T_{m}^{m}(\omega-1)\bigg[\alpha (p+1)-(p-\alpha)(p-1)\bigg]\Bigg) %
-\frac{\omega }{\phi ^{2}}(p-\alpha )\phi^{\prime }\cdot \phi^{\prime }%
\Bigg\}\right. ,  \label{B6}
\end{eqnarray}
with the particular assumption that only the bulk respect the Brans-Dicke
gravity, disregarding, then, $\nabla_\mu\phi$ terms. This assumption is
quite conceivable, since it does not jeopardize the standard four
dimensional gravity. Now, inserting the energy-momentum partial traces and
integrating over the internal space, we have 
\begin{eqnarray}
\oint W^{\alpha +1}\Bigg\{ \alpha W^{-2}\overset{\_}{R}&+&\left.(p-\alpha )%
\tilde{R}+\frac{8\pi }{\phi }\frac{1}{[(D-1)+(D-2)\omega ]}\Bigg(-\dfrac{1}{2%
}\Phi^{\prime }\cdot\Phi^{\prime }\lbrack A(p+1)+B(D-p-3)]\right.  \notag \\
&-&\left. V(\Phi )[A(p+1)+B(D-p-1)]\Bigg)-\dfrac{\omega }{\phi ^{2}}%
(p-\alpha )\phi^{\prime }\cdot\phi^{\prime }\Bigg\}=0,\right.  \label{b10}
\end{eqnarray}%
where%
\begin{eqnarray}
A &\equiv &(D-p-2)(p-2\alpha)-\alpha (D-p-3)\omega+(p-\alpha )(D-p-1)\omega ,
\notag \\
B &\equiv &(p+1)(\omega+1)(2\alpha-p),  \label{b11}
\end{eqnarray}%
are constant parameters.

We are now in position of particularize the above formalism to the
five-dimensional bulk case. Then, setting $D=5$ and $p=3$ we find 
\begin{eqnarray}
\oint W^{\alpha+1}\Bigg\{-\frac{8\pi}{\phi}\frac{1}{4+3\omega}\Bigg(\frac{1}{%
2}\Phi^{\prime }\cdot\Phi^{\prime
}[12(1&+&\left.\omega)-8\alpha-6\omega(\alpha-1)]+V(\Phi)[12(-1+\omega)%
\right.  \notag \\
&+&\left.8\alpha+6\omega(\alpha-1)]\Bigg)-\frac{\omega}{\phi^2}%
(3-\alpha)\phi^{\prime }\cdot\phi^{\prime }\Bigg\}=0.\right.
\end{eqnarray}
There are at least two choices for the parameter $\alpha$ bringing relevant
information about the system. To begin with, notice that for $\alpha=-1$ we
have 
\begin{equation}
\dfrac{(5+6\omega )}{4+3\omega }\oint \dfrac{1}{\phi }\Phi ^{\prime }\cdot
\Phi ^{\prime }=\dfrac{10}{4+3\omega }\oint \dfrac{1}{\phi }V(\Phi )+\dfrac{%
\omega }{4\pi }\oint \dfrac{1}{\phi ^{2}}\phi^{\prime }\cdot \phi^{\prime
}=0,  \label{b20}
\end{equation}
which in the limit $\omega\rightarrow\infty$ $(\phi\rightarrow cte)$ leads,
as expected, to the same consistency condition given by Eq. (\ref{no-go}),
expliciting the no-go theorem. Returning to Eq. (\ref{b20}), it is remarkable
that we have a relaxing of the consistency condition, in the same spirit of
Eq. (\ref{frelevant}). In the present case, however, we see an odd
characteristic coming exclusively from the Brans-Dicke theory, namely, the
presence of the scalar field potential in the consistency condition.
Whenever the choice $\alpha=-1$ is made, it eliminates the potential of this
consistency condition in the usual (General Relativity) and $f(R)$
framework. The result encoded in Eq. (\ref{b20}) shows that within the
Brans-Dicke theory the potential used to set the scalar field shape is,
indeed, relevant for the obtainment of a well defined model. It is also
interesting the choice $\alpha=3$, since the $\phi^{\prime
}\cdot\phi^{\prime }$ contribution vanishes in this case, leading to 
\begin{equation}
\oint \frac{W^4}{\phi}\Phi^{\prime }\cdot\Phi^{\prime }=(4\omega+2)\oint V,
\label{ult}
\end{equation}
expliciting, once again, the relevance of the Brans-Dicke theory for the
mitigation of the constraints coming from the braneworld sum rules.

\section{Concluding Remarks}

The necessity of a braneworld scenario composed by smooth branes is widely
accepted from the physical point of view. Indeed, the very fact that
standard gravitational theories should be replaced above the Planck lenght
scale points to the need of branes with some thickness. On the other hand,
the outcome obtained from the formulation of the braneworld sum rules is
exhaustive: within General Relativity, smooth branes are not allowed in five
dimensions (assuming a compact extra dimension).

Revisiting the sum rules, this time with the auspices of two modified
gravitational theories, $f(R)$ and Brans-Dicke, we see that in both cases it
is possible to relax the consistency conditions and smooth branes are
possible, in principle. 

Our framework is mainly based upon that fact that the integral 
of a total divergence over a compact internal space must be equal to zero by means of the orbifold symmetry. By compact, 
then, we mean a finite internal space without boundaries. Particularly, this approach sounds reasonable 
in five dimensions since in this case the internal space is nothing but the extra dimension. 

In the particular Brans-Dicke framework, it is also interesting to note that the scalar field potential can contribute
to the sum rule allowing for the smooth brane. This is exclusive from the
formulation within the Brans-Dicke theory and naturally opens up new possibilities of
smooth brain modeling, enriching scenarios constructed in such framework. This possibility, raised in the realm of the Brans-Dicke theory, may be faced as a bonus in braneworld modeling. As we have mentioned, in General Relativity, smooth branes are not allowed in five dimension regardless the scalar field potential. Regardless the model, for short. In this vein, there are plenty of models, respecting the constraint (\ref{ult}) for instance, with can be implemented within Brans-Dicke gravity. We believe that this possibility may encourage model builders outside the General Relativity framework.

\section*{Acknowledgments}

ASD and JMHS thanks to CNPq for financial support. GPB thanks to FAPESP for
financial support, and PMLTS acknowledge CAPES for financial support.

\end{document}